\documentclass[useAMS,usenatbib]{./mn2e}
\usepackage{./psfig}

\newcommand{\Wo}{\mbox{${\rm W}_0$}}
\newcommand{\msun}{\mbox{${\rm M}_\odot$}}

\newcommand{\rsun}{\mbox{${\rm R}_\odot$}}
\newcommand{\kms}{\mbox{${\rm km~s}^{-1}$}}

\newcommand{\nbody}{\mbox{{{\em N}-body}}}
\newcommand{\thm}{\mbox{${t_{\rm hm}}$}}

\newcommand{\thc}{\mbox{${t_{\rm hm}}$}}

\newcommand{\trlx}{\mbox{${t_{\rm rlx}}$}}

\newcommand{\mm}{\mbox{$\langle m \rangle$}}

\newcommand{\rcore}{\mbox{${r_{\rm core}}$}}
\newcommand{\rc}{\mbox{${r_{\rm core}}$}}

\newcommand{\rv}{\mbox{${r_{\rm vir}}$}}
\newcommand{\rhm}{\mbox{${r_{\rm hm}}$}}

\newcommand{\rt}{\mbox{${r_{\rm tide}}$}}

\newcommand{\rJ}{\mbox{${r_{\rm Jacob}}$}}

\newcommand{\vesc}{\mbox{$v_{\rm esc}$}}

\newcommand{\rlof}{\mbox{RLOF}}

\newcommand{\nbsbag}{\mbox{$N_{\rm bss:Be:gs}$}}
\newcommand{\fsbt}{\mbox{$f_{\rm s:b:t}$}}

\def\apgt{\ {\raise-.5ex\hbox{$\buildrel>\over\sim$}}\ }
\def\aplt{\ {\raise-.5ex\hbox{$\buildrel<\over\sim$}}\ }

\title[Star cluster ecology V]
      {Star cluster ecology V: Dissection of an open star cluster---spectroscopy}

\author[Simon F.\ Portegies Zwart et al.]
        {
	Simon F.\ Portegies Zwart,$^{1}$\thanks{E-mail:
	spz@science.uva.nl; KNAW fellow} 
    	Piet Hut,$^{2}$
	Stephen L.\ W.\ McMillan$^{3}$
	and
	Junichiro Makino,$^{4}$\\
$^1$    Astronomical Institute 'Anton Pannekoek',
	University of Amsterdam, Kruislaan 403, 
	1098SJ Amsterdam, the Netherlands, \\
	Section Computational Science,
	University of Amsterdam, Kruislaan 403, 
	1098SJ Amsterdam, the Netherlands \\
$^2$		Institute for Advanced Study,
		Princeton, NJ 08540, USA \\
$^3$ 	Department of Physics,
		Drexel University,
                Philadelphia, PA 19104, USA \\
$^4$ 	Department of Information Science and Graphics,
		College of Arts and Science,
		University of Tokyo, 3-8-1 Komaba,
		Meguro-ku, Tokyo 153, Japan
        }

\begin{document}

\date{Accepted 2145 December 32. Received 1687 December -1; in
original form 1900 October 11}

\maketitle
\pagerange{\pageref{firstpage}--\pageref{lastpage}} \pubyear{2002}

\maketitle

\label{firstpage}

\begin{abstract}
We have modeled in detail the evolution of rich open star clusters
such as the Pleiades, Praesepe and Hyades, using simulations that
include stellar dynamics as well as the effects of stellar evolution.
The dynamics is modeled via direct {\nbody} integration, while the
evolution of single stars and binaries is followed through the use of
fitting formulae and recipes.  The feedback of stellar and binary
evolution on the dynamical evolution of the stellar system is taken
into account self-consistently.

Our model clusters dissolve in the tidal field of the Galaxy in a time
span on the order of a billion years.  The rate of mass loss is rather
constant, $1\sim2\,\msun$ per million years.  The binary fraction at
first is nearly constant in time, then increases slowly near the end
of a cluster's lifetime.  For clusters which are more than about
$10^8$\,years old the fractions of stars in the form of binaries,
giants and collision products in the inner few core radii are
considerably higher than in the outer regions, beyond the cluster's
half mass radius.  When stars with masses $\apgt 2$\,\msun\, escape
from the cluster, they tend to do so with velocities higher than
average.

The stellar merger rate in our models is roughly one per 30 million
years.  Most mergers are the result of unstable mass transfer in close
binaries ($\sim 70$\%), but a significant minority are caused by
direct encounters between single and binary stars.  While most
collisions occur within the cluster core, even beyond the half mass
radius collisions occasionally take place.  We notice a significant
birthrate of X-ray binaries, most containing a white dwarf as the
donor.  We also find some X-ray binaries with a neutron-star donor,
but they are relatively rare.  The persistent triple and higher order
systems formed in our models by dynamical encounters between binaries
and single stars are not representative for the multiple systems
observed in the Galactic disk or in open clusters.  We conclude that
the majority of multiples in the disk probably formed when the stars
were born, rather than through later dynamical interactions.

\end{abstract}



\section{Introduction}

\begin{table*}
\caption[]{Observed and derived parameters for several open star
clusters with which our simulations may be compared.  References to
the literature (second column) are:
(a) Pinfield et al.\, (1998);\nocite{1998MNRAS.299..955P}
    Raboud \& Mermilliod, (1998);\nocite{1998A&A...329..101R}
    Bouvier et al (1998);\nocite{1997A&A...323..139B}
(b) Abt \& Levy (1972)\nocite{1972ApJ...172..355A}; 
    Dachs, J \& Kabus (1989)\nocite{1989A&AS...78...25D};
    Hawley et al. (1999).\nocite{1999AJ....117.1341H}
(c) Harris et al. (1993);\nocite{1993AJ....106.1533H}
    Ianna et al (1987);\nocite{1987AJ.....93..347I}
    Cox (1954).\nocite{1954ApJ...119..188C}
(d) Andrievsky (1998);\nocite{1998A&A...334..139A} 
    Jones \& Stauffer (1991);\nocite{1991AJ....102.1080J} 
    Mermilliod \& Mayor (1999);\nocite{1999astro.ph.11405M}
    Mermilliod et al. (1990);\nocite{1990A&A...235..114M}
    Hodgkin et al. (1999).\nocite{1999MNRAS.310...87H}	
(e) Perryman et al.\, (1998 and references therein)
    \nocite{1998A&A...331...81P} 
    Reid \& Hawley (1999);\nocite{1999AJ....117..343R}
%
(f) Frandsen et al. (1989);\nocite{1989A&A...215..287F} 
    Hartwick \& Hesser (1971);\nocite{1971PASP...83...53H} 
    Sandrelli et al. (1999).\nocite{1999MNRAS.309..739S}
(g) Hawley et al. 1999 \nocite{1999AJ....117.1341H}; 
    Nordstr\"om et al. (1997);\nocite{1997A&A...322..460N} 
    Nordstr\"om et al.\,(1996),\nocite{1996A&AS..118..407N}
Subsequent columns give (3) the distance to the cluster (in pc), (4)
the cluster age (in Myr), (5) the half mass relaxation time, (6) the
crossing time, (7) the total mass (in \msun), (8) estimate
for the half mass radius (in pc), and (9) the core radius.  In cases
where the relaxation time is not given in the literature, we
calculate it (see Paper IV); these entries
are indicated by $^\star$.  The final two columns contain information
about cluster stellar content.  The column labeled $\fsbt$ indicates
the number of single stars, binaries and triples (separated by
colons).  For clusters where the numbers are given directly by
observations, the table gives the observed numbers of each system.  If
the binary fraction is derived by other methods, we give the relative
fractions normalized to the number of single stars.  The last column
($\nbsbag$) gives the number of observe blue stragglers, Be stars and
giants, separated by a column.}
\bigskip
\begin{tabular}{llrrrrrrrrrcc} \hline
name    &ref.&$d$ &$t$&    \trlx&\thm&$M$ &\rt\  &\rhm\   &\rc\ 
                                                      &\fsbt&\nbsbag \\
   &  &[pc]&\multicolumn{3}{c}{--- [Myr] ---}
      &[\msun]&\multicolumn{3}{c}{--- [pc] ---}\\
(1)&(2)&(3)&(4)&(5)&(6)&(7)&(8)&(9)&(10)&(11)&(12)\\
\hline
Pleiades &a  & 135&  115    &     150& 8    &$\sim$1500&   16& 2--4 & 1.4  
                                                     &137:60:2 & 3:3:?\\
NGC2516 &b  & 373&  110     &{\sl 220}&  &{\sl 1000}&     & 2.9 & &16:6:?&6:3:4\\
NGC2287 &c  & 675&  160--200&  ---    &  &          & 6.3 &
						&   &1:0.6:? &3:1:8\\
Praesepe&d  & 174&  400--900&{\sl 220}&  &1160&     & 12  &  2.8  
                                                      &1:0.3:0.03 &?:5:?\\
Hyades  &e  &  46& 625      &{\sl 320}&15&1027& 10.3& 3.7$^1$&2.6
                                                      &1:0.4:0 & 1:0:4\\
NGC 2660&f  & 2884&900--1200&{\sl 260}&  & 400&  ?  & 4 &$\sim 1$
                                                      &0:0.3:? & 18:4:39\\
NGC 3680&g  & 735&1450      &{\sl 28} &  & 100&  4.3& 1.2     &$\sim1$
                                                      &44:25:0 & 4:4:17\\
\hline
\end{tabular}

\label{Tab:observed} 
\end{table*}

\begin{table*}
\caption[]{Initial conditions and parameters for the selected models.
The first column gives the model name, followed by the cluster mass
(in \msun), the King (1966) parameter $W_0$, the distance to the
Galactic center (in kpc), the initial relaxation time and half mass
crossing time (both in Myr). The remaining columns give the location
of the cluster's Jacobi surface along the $X-$ (towards the Galactic
center), $Y-$ and $Z-$ (towards the Galactic pole) axes, and the
initial virial radius, half mass radius and core radius (all in
parsec).}
\begin{flushleft}
\begin{tabular}{l|rrrrrrrrrrr} \hline
name &$M$    &\Wo&$R_{\rm Gal}$
             & \trlx &\thc&\multicolumn{3}{c}{\rJ} &\rv &\rhm &\rc \\
     &[\msun]&&[kpc]&\multicolumn{2}{c}{[Myr]}
              &\multicolumn{3}{c}{[pc]}  & \multicolumn{3}{c}{[pc]} \\ \hline
W4   & 1708& 4& 6.8&135&4.07&  14.5& 9.7& 7.2   & 2.5&2.14 & 0.83 \\
W6   & 1603& 6&10.4&140&4.15&  21.6&14.4&10.8   & 2.5&2.00 & 0.59 \\ \hline
\end{tabular}
\end{flushleft}
\label{Tab:init} 
\end{table*}

Open clusters are useful laboratories for studying the interplay
between single star evolution, binary star evolution and stellar
dynamics.  Unlike their bigger (and older) siblings, the globular
clusters, they contain a manageable number of stars, and the evolution
of the majority of the stars is not expected to be strongly affected
by the dynamics.  Still, a significant number of collisions and
subsequent stellar mergers can take place, as well as dynamically
induced exchange reactions between single stars and binaries.  For all
these reasons, the usual zoo of objects created in binary stellar
evolution is significantly enlarged by the presence of even more
exotic specimens that could not have formed {\it in vitro} through
isolated evolution, but only {\it in vivo} through the dynamical
interplay of initially unrelated (single or multiple) stars.

The simulations reported here have been run on a GRAPE-4 (Makino et
al.\ 1997)\nocite{1997ApJ...480..432M} system, a special-purpose
computer designed to speed up stellar-dynamical calculations.  While
the models in this paper contain only $\sim3$,000 stars, appropriate
for open clusters, we have started to use the next-generation
special-purpose computer, the GRAPE-6, to extend our simulations to
include one or two orders of magnitude more stars.  This will enable
us to model globular clusters on a star-by-star basis, including those
with very dense cores where most stars are influenced by various
`traffic accidents' in the form of encounters and mergers.  The
simulations reported here thus play a double role: astrophysically,
they provide insight in the evolution of open clusters, and
computationally, they help pave the way for the modeling of the more
complex globular clusters.

In this paper we describe a series of self-consistent simulations of
the evolution of open star clusters.  All important effects are
included to some degree of realism: stellar evolution, binary
evolution, the internal dynamical evolution and the effects of the
tidal field of the Galaxy.  This is the fifth paper in a series in
which we have gradually increased the `ecological' complexity of
stellar interactions to a realistic level.  In this paper we analyze
the results of the same eight calculations performed for our previous
Paper IV (Portegies Zwart et al.\ 1999)\nocite{pzmmh99}. There we
concentrated on global cluster properties and compared them with
observed open clusters, such as the Pleiades, Praesepe and
Hyades. Specifically, we took a photometric standpoint and studied
changes in the Hertzsprung-Russell diagram and cluster morphology as
functions of time and initial conditions.  Here we concentrate on the
binary population and on higher-order multiple systems, and we study
the dynamical and observational effects of these binaries and
multiples on the evolution of the stellar system as a whole.

Detailed descriptions of the numerical methods used and global
assumptions made in our calculations are given in Paper IV.  Since we
analyze the same data from a different perspective we do not repeat
this information here.  Instead, we quickly review the initial
conditions of our models, and restate the methods used; for more
details, see Paper IV.

\section{Methods}
In this section we briefly discuss the initial conditions of our
models and the numerical techniques used in our model calculations.
For details, refer to Appendix B and C of Paper IV (Portegies Zwart et
al.\ 1999)\nocite{pzmmh99}.

\subsection{Initial conditions}

We are interested in moderately rich ($\sim 1000$\, stars) open
clusters of intermediate age ($\aplt 1$\,Gyr).  Starting from the
currently observed mass and dynamical properties of such clusters we
have reconstructed two plausible sets of initial conditions, which are
presented in Table\,\ref{Tab:init}. Each calculation is performed four
times with different random seeds in order to improve statistics. The
notation in this paper is identical to that used in Paper IV (see its
Appendix A for an overview).

For our simulations we assume a Scalo (1986)\nocite{scalo86} initial
mass function, with minimum and maximum masses of 0.1\,\msun\, and
100\,\msun, respectively, and mean mass $\mm\simeq 0.6\msun$ at birth.
Consistent with the above mass estimates, our simulations are
performed with 1024 single stars and 1024 binaries, for a total of
3096 (3k) stars.

Stars and binaries within our model are initialized as follows.  A
total of 2k single stars are selected from the initial mass function
and placed in an equilibrium configuration in the selected density
distribution (see below).  We then randomly select 1k stars and add a
second companion star to them.  The masses of the companions are
selected randomly from the IMF between 0.1\,\msun\, and the primary
mass (see Tab.\,\ref{Tab:Binit}).  For the adopted binary fraction,
the restricted secondary mass range translates into an overall mean
mass of $0.53\msun$.  Then the other binary parameters
are determined.  Binary eccentricities are selected from a thermal
distribution between 0 and 1.  Orbital separations $a$ are selected
with uniform probability in $\log a$ between Roche-lobe contact and an
upper limit of 5,000 A.U. (= $10^6$\,\rsun, or 0.02\,pc).  When a
binary appears to be in contact at pericenter, that particular orbit
choice is rejected and new orbital parameters are selected.
Table\,\ref{Tab:Binit} gives an overview of the various distribution
functions from which stars and binaries are initialized.  (Figure
\ref{fig:Porb_ecc} shows the initial distributions of orbital period
and eccentricity.)

\begin{table}
\caption[]{Initial conditions for the stellar and binary population.
The first column gives the parameter, the second column gives the
functional dependence, followed by the lower and upper limits adopted.
(\rlof\, indicates that the initial binary may be semi-detached.)}
\begin{tabular}{ll|rr} \hline
parameter          & function               & lower & upper \\ 
                   &                        & \multicolumn{2}{c}{limits} \\
mass function      & Scalo                  & 0.1\,\msun  & 100\,\msun \\
secondary mass     & $P(m) = {\rm constant}$& 0.1\,\msun  & $M_{\rm prim}/\msun$ \\
orbital separation & $P(a) = 1/a$           & \rlof       & 5000\,AU \\
eccentricity       & $P(e) = 2e$            & 0           & 1    \\ \hline
\end{tabular}
\label{Tab:Binit} \end{table}

We select initial density profiles from the density distributions
described by Heggie \& Ramamani (1995)\nocite{1995MNRAS.272..317H}
with $\Wo=4$ and $\Wo=6$, and refer to these models as W4 and W6,
respectively, throughout this paper.  We have performed four
independent simulations for each set of initial conditions,
labeled $I$ to $IV$.

All our cluster models start with the same virial radius of $R_0 =
2.5$ pc.  This implies a conveniently constant scaling between the
cluster dynamical time scale $\sim\left(GM_0/R_0^3\right)^{-1/2}$ ($=
1.5$ Myr for $M_0 = 1600 \msun$) and the time scale for stellar
evolution.  Each cluster is assumed to precisely fill its Jacobi
surface at birth.  Given the observed Oort constants in the solar
neighborhood, we find that a model with $\Wo=6$ has to be placed
slightly farther (10.4\,kpc) from the Galactic center than the Sun,
while a model with $\Wo=4$ has to reside somewhat closer (6.8\,kpc),
in order to obey the above relationships.

For a total cluster mass of 1600\,\msun, the Lagrange points of our
two standard clusters lie at distances 14.5 pc ($\Wo=4$) and 21.6 pc
($\Wo=6$), from the cluster center.  Stars are removed from a
simulation when their distance to the cluster's density center exceeds
twice the distance to the first Lagrangian point.
Table\,\ref{Tab:init} reviews the selected parameters and initial
conditions.

\subsection{The \nbody\ integrator}\label{Sect:integrator}
The \nbody\ integration in our simulations is carried out using the
{\tt kira} integrator, operating within the Starlab software
environment (McMillan \& Hut 1996; Portegies Zwart et al.\
1999).\nocite{1996ApJ...467..348M}\nocite{pzmmh99} {\tt Kira} uses a
fourth-order Hermite scheme (Makino \& Aarseth
1992),\nocite{1992PASJ...44..141M} incorporates block time steps
(McMillan 1986a; 1986b; Makino
1991),\nocite{1986ApJ...307..126M}\nocite{1986ApJ...306..552M}
\nocite{1991ApJ...369..200M} includes special treatments of close
two-body and multiple encounters of arbitrary complexity, and a robust
treatment of stellar and binary evolution and stellar collisions (see
below).  As mentioned above, the special-purpose GRAPE-4 (Makino et
al.\ 1997)\nocite{1997ApJ...480..432M} system is used to accelerate
the computation of gravitational forces between stars.

\subsection{Stellar evolution}\label{Sect:SeBa}
The evolution of the stars is adapted from the prescription by
Portegies Zwart \& Verbunt (1996, \S 2.1).\nocite{pzv96} However,
some changes are made to the mass loss in the main-sequence stage for
massive stars and for mass-transfer remnants. We follow the details of
model B from Portegies Zwart \& Yungelson
(1998).\nocite{1998A&A...332..173P} For our treatment of stellar mass
loss, see Portegies Zwart et al.\ (1998).\nocite{1998A&A...337..363P}

Neutron stars receive a high-velocity kick at the moment of their
formation. The reasons for the occurrence of these high kick
velocities have been discussed in detail by Portegies Zwart \& van den
Heuvel (1999)\nocite{1999NewA....4..355P} and the implications for
binary evolution are discussed by Portegies Zwart
(2000).\nocite{2000astro.ph.5021PZ} The distribution from which the
kick velocity should be taken is less certain. We adopt the velocity
distribution proposed by Harmann (1997)\nocite{1997A&A...322..127H}
which has a dispersion velocity of 450\,\kms.  Each neutron star
formed received a kick chosen from this distribution and oriented in a
random direction.

\section{Results}

\subsection{Overall evolution of the models}\label{sect:evolution}

In this section we review the global properties of models W4 and W6,
summarizing the more extensive discussion presented in Paper IV.
Where we discuss individual models, we focus on the same two models
discussed in Paper IV (model W4-IV and W6-III).  In other cases the
data for several models, or even all models, are combined in order to
improve statistics; those cases are indicated specifically below.

\begin{figure}
\hspace*{1.cm}
\psfig{figure=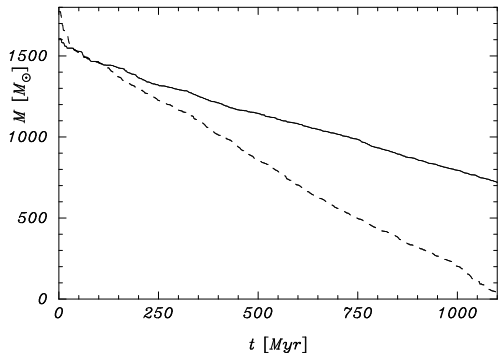,width=7.5cm,angle=-90} 
\caption[]{Total mass as a function of time for model W4 (dashes) and
W6 (solid).}
\label{fig:W46_tM}
\end{figure}

\begin{figure}
\hspace*{1.cm}
\psfig{figure=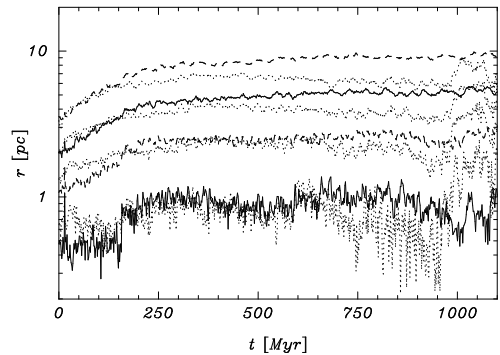,width=7.5cm,angle=-90} 
\caption[]{Evolution of the 5\%, 25\%, 50\% and 75\% Lagrangian radii
for model W6 (solid lines for the 5\% and 50\%, dashed lines for the
25\% and 75\% Lagrangian radii). Lagrangian radii for model W4 are
presented as dotted lines.  }
\label{fig:W46_trL}
\end{figure}

Figures \ref{fig:W46_tM} and \ref{fig:W46_trL} present the overall
evolution for our two sets of cluster initial conditions.  Figure
\ref{fig:W46_tM} shows the time evolution of the cluster mass.  Models
W4 and W6 lose mass at roughly constant rates of about 1.4\,\msun\,
(0.09\%) and 0.82\,\msun\, (0.05\%) per million years, respectively.
These mass loss rates result in the disruption of the cluster at about
1 Gyr for model W4 and around 2\,Gyr for the more concentrated model
W6.  The higher mass loss rate of model W4 is mainly attributable to
its closer proximity to the Galactic center and not per se to its
lower concentration.  These rates are consistent with a mass loss rate
of 1.2\,{\msun} per million years for the slightly more massive star
cluster studied by Hurley et al. (2001;
2002).\nocite{2001MNRAS.323..630H}\nocite{2002MNRAS.329..897H} The
mass-loss rates per half mass relaxation time derived by Portegies
Zwart et al.\,(2001)\nocite{2001ApJ...546L.101P} for dense star
clusters near the Galactic center are about a factor of four higher
than the corresponding rates for the clusters studied here.

Figure \ref{fig:W46_trL} shows the evolution of the Lagrangian radii
containing 5\%, 25\%, 50\% and 75\% of the cluster mass, for models W4
and W6.  Both clusters expand by about a factor of three during the
first half-mass relaxation time ($\sim 140$\,Myr).  It is not clear
whether the clusters experience any significant core collapse during
this early phase. In Paper IV we concluded that the clusters do not
experience core collapse due to the effects of binary activity, which
tends to counteract core contraction (see McMillan et al.\ 1990,
1991)\nocite{1990ApJ...362..522M}\nocite{1991ApJ...372..111M}.  At
later times ($t\apgt 650$\,Myr for model W4 and somewhat later for
model W6), the Lagrangian radii decrease again as the cluster starts
to dissolves in the tidal field of the Galaxy.

\subsection{Global binary properties}\label{Sect:properties}

Both models start with 50\% binaries, so two-thirds of the stars are
initially members of binary systems.  The adopted upper limit
($10^6$\,\rsun) on the initial semi-major axis means that most initial
binaries are relatively wide. The fraction of hard ($|E| \apgt1 kT$)
binaries was $0.46\pm0.01$ for all models, irrespective of the initial
density profile. (The thermodynamic unit $kT$ is defined in Paper $IV$).

\begin{figure}
\psfig{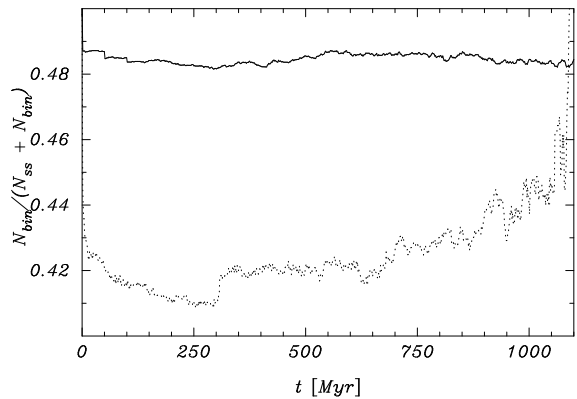} 
\caption[]{ Binary fractions as a function of time for models W4
(dotted line) and W6 (solid line).  }
\label{fig:binfrac_W46}
\end{figure}

Figure \ref{fig:binfrac_W46} shows the evolution of the binary
fraction.  Shortly after the start of the runs, the binary fractions
drop to about 42\% and 48\% for models W4 and W6, respectively, and
remain roughly constant thereafter.  This initial decrease in binary
frequency is consistent with the disruption of all soft ($|E| \aplt1
kT$) binaries.  Supernova explosions dissociated 1.8\% of the binaries
in the W4 models, and about half that number in the W6 models.  This
difference is mainly the result of random fluctuations.  Fluctuations
in the binary fraction after $\sim 100$\,Myr are mainly the result of
escapers and mergers.  The number of mergers resulting from unstable
mass transfer or direct collisions is $12.1\pm0.5$\% for models W4 and
W6.  The escape rate of binaries closely follows the escape rate of
single stars; both are relatively constant with time.  Model W6 loses
one binary per 2Myr; the escape rate for model W4 is about twice as
high.

\subsection{Evolution of binary parameters}

The ecological interplay between stellar (and binary) evolution,
stellar (and binary) dynamics, and the external tidal influence of the
Galaxy transforms the initial distributions of stars and binaries in
complex ways.  We assume that initially all binaries are distributed
throughout the cluster in the same way as single stars.  However,
since binaries are on average more massive than single stars, they
subsequently tend to sink to the cluster center, where superelastic
encounters quickly modify the spatial distributions of both stars and
binaries.

These encounters come in many forms, from the more common
single-star--binary and binary--binary types to the rare forms that
involve triples and occasionally even higher-order multiple systems.
The combined effects of such encounters significantly modifies the
binary distribution functions, in terms of radius and binding energy
as well as stellar type.  In addition, stellar evolution and isolated
binary evolution also cause binary parameters to evolve in time,
independent of the occurrence of dynamical close encounters.
In this subsection, we will present some of the net results of these
complex processes.

\subsubsection{Distributions of orbital period and eccentricity}

\begin{figure}
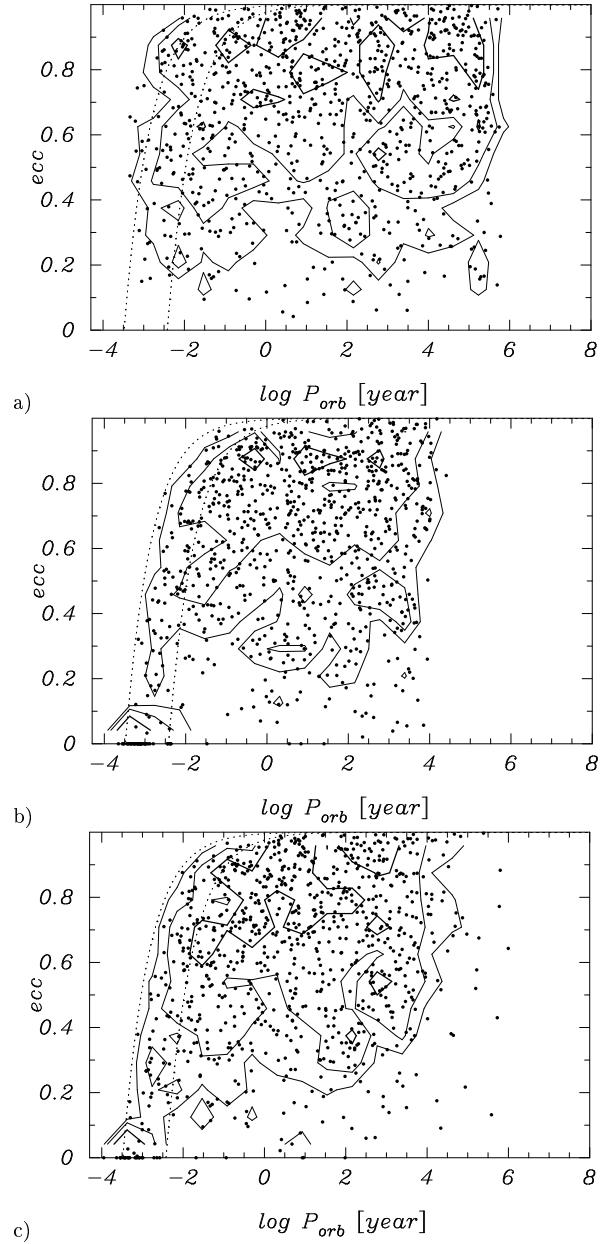

a)\psfig{figure=./fig_W6t0_Pe_N871.ps,width=7.5cm,angle=-90} 
b)\psfig{figure=./fig_W4t600_Pe.ps,width=7.5cm,angle=-90} 
c)\psfig{figure=./fig_W6t600_Pe_N871.ps,width=7.5cm,angle=-90} 
\caption[]{Scatter plots for the orbital period versus the
eccentricity for 871 binaries for various models and moments in time:
a) initial distribution, b) the distribution for model W4 at an age of
600\,Myr, c) model W6 at $t=600$\,Myr.  Each panel is constructed
using 871 binaries.  Contours give the 0.5, 0.25 and 0.125
iso-density curves of the distribution in the two coordinates plotted
here. The left dotted curve indicates at which orbital period
and eccentricity a binary with a 0.1\,\msun\, zero-age primary
circularizes during the time span of our simulations. The right
dotted curve is for a 2.26\,\msun\, primary at zero age.  }
\label{fig:Porb_ecc}
\end{figure}

Figure \ref{fig:Porb_ecc}a shows the initial distribution of orbital
period (in years) and eccentricity of a subset of the primordial
binaries of model W6 (the initial conditions for model W4 are
identical).  Figures \ref{fig:Porb_ecc}b and c show the same
distributions for models W4 (panel b) and W6 (panel c) at 600\,Myr.
In the middle panel, for model W4, only 871 binaries remained.  In
order to facilitate visual comparison, the upper and lower panels also
show only 871 binaries out of the larger numbers that could have been
plotted.

A striking feature of the three panels in Fig\,\ref{fig:Porb_ecc}, is
the lack of wide ($p_{\rm orb} \apgt 10^4$\,yr) binaries at later
times.  These binaries are completely absent in model W4, while in
model W6 a small population of wide binaries remains.  This deficiency
in wide binaries in evolved clusters is caused by ionization of the
soft binaries by dynamical encounters (see also Figure
\ref{fig:binfrac_W46}). For model W4 this process is more efficient
than for the more concentrated models because 1) the W4 models
experience a high density phase during their early evolution, and 2)
the W6 models have extended outer regions with relatively low stellar
densities because of the weaker tidal field.  All surviving binaries
with $p_{\rm orb} \apgt 10^5$\,year are located well beyond the
cluster's half mass radius, and most contain at least one white dwarf.
In many cases, the orbital periods of these non-interacting binaries
has increased substantially due to the large amount of mass lost from
one (or both) of the component stars.

Binaries located above and to the left of the left-most dotted curves
experience strong tidal effects during their early evolution.  As a
result, they quickly circularize, as can be seen by their
precipitation to the zero-eccentricity line at the bottom in the last
two panels.  The majority of these systems contain at least one white
dwarf or helium star.  Occasionally a binary may enter the 'forbidden'
left-most area at a later time as a result of a strong dynamical
encounter. Such a binary will either be quickly circularized or its
components will collide and merge to form a single star (see
\S\ref{Sect:Collisions}).

\begin{figure}
\psfig{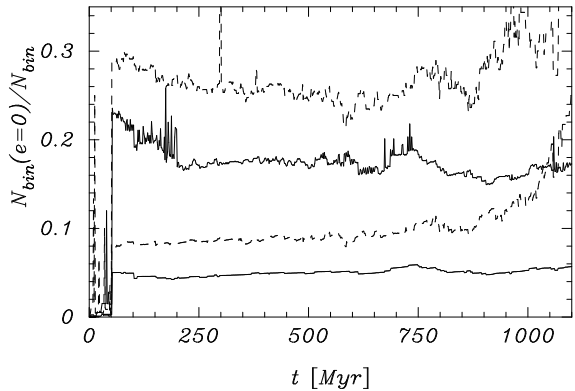} 
\caption[]{ The relative number of binaries that are tidally
circularized, as a function of time, for the models W4 (dashed lines)
and W6 (solid lines).  In each case, the lower lines present the overall
fraction of binaries that are circularized, while the top curves
indicate the fraction of circularized binaries among those binaries
that are very hard ($|E_{\rm bin}| \apgt 1000 kT$). }
\label{fig:Bcircfrac}
\end{figure}

Figure \ref{fig:Bcircfrac} shows how the percentage of tidally
circularized binaries evolves over time.  Not surprisingly, the very
hard binaries, which have high binding energy and therefore tight
orbits, have on average undergone much stronger tidal interactions and
therefore have had a much larger chance to circularize.  The four
percentages presented in the figure---hard binaries and all binaries
for each of the two classes of model---at first change little during
the evolution of the cluster, but they increase at later times in the
case of model W4.

Note that in Fig\,\ref{fig:Porb_ecc} there are some circularized
binaries with long orbital periods, far to the left of the right-most
dotted line, in the middle and lower panel.  These are absent in the
upper panel, which shows the initial conditions.  The tidal
circularization of these systems occurred when one of the stars
ascended the giant branch before the onset of mass transfer.  While in
many cases this type of evolution leads to a later shrinking of the
orbit, in some rare cases the binaries remain wide after a period
of modest mass transfer.

\begin{figure}
\hspace*{1.cm}
\psfig{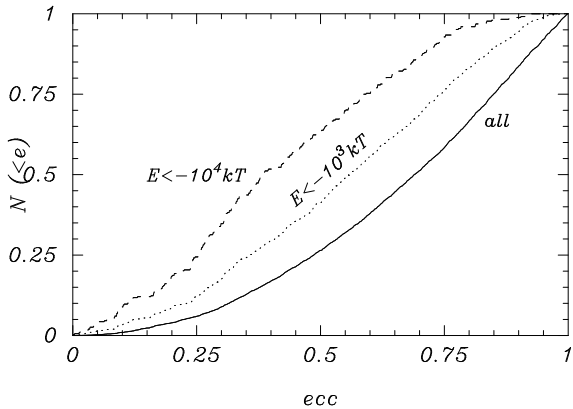} 
\caption[]{Cumulative distribution of the binary eccentricity. Solid
line, initial conditions (the thermal distribution). The dashed and
dotted line represent the eccentricity distributions at an age of
600\,Myr of binaries with a binding energy $E_{\rm bin} < -10^3 kT$
and $E_{\rm bin} < -10^4 kT$, respectively.  For the latter two curves
the data for all models W4 and W6 are combined. However, we excluded
those binaries which have been circularized by tidal forces. }
\label{fig:W46_cum_ecc}
\end{figure}

Since tidal forces drop off quickly with distance, in most cases they
either succeed in fully circularizing a binary, or they do not cause
much of an effect.  Consequently, the eccentricity distribution of
most non-circularized binaries changes little with time.  Figure
\ref{fig:W46_cum_ecc} illustrates this by presenting the cumulative
distribution of all binaries with the exception of those with zero
eccentricity.  Note that this general argument does not hold for the
hardest binaries: when we select only the
hard ($|E_{\rm bin}| \apgt 1000 kT$) or super-hard ($|E_{\rm bin}|
\apgt 10^4 kT$) binaries, clear deviations from the initial thermal
distribution are evident.  This is not surprising: a $10^4 kT$ binary
simply does not have enough room to develop an eccentricity of, say,
0.9, because the semimajor axis is too small.

\subsubsection{Radial distribution of binaries}

\begin{figure}
\hspace*{1.cm}
\psfig{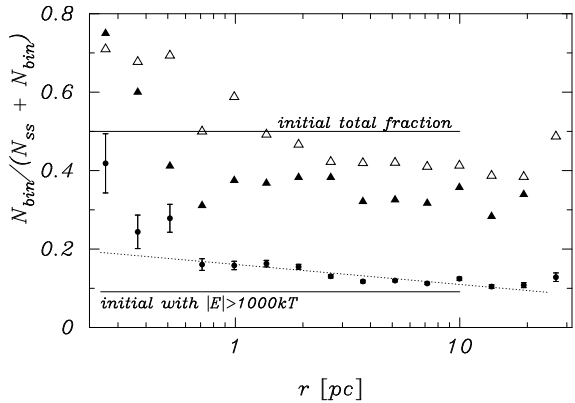} 
\caption[]{ 
Fraction of binaries $N_{\rm bin}/(N_{\rm ss}+N_{\rm bin})$ as
function of the distance to the cluster center.
The upper solid horizontal line shows the initial binary fraction over
the entire star cluster. The cut off at about 10\,pc is close to the
location of the tidal radius. (Initially there were no stars beyond the
tidal radius.)
The lower solid curve presents the initial distribution for binaries
with $|E_{\rm bin}| > 1000 kT$.  
The open triangles give the distribution of all binaries for model W6
at an age of 600\,Myr. The solid triangles present the distribution of
all binaries for model W4 at an age of 600\,Myr.
The bullets (with $1\,\sigma$ Poissonian error bars) present the
fraction of hard $|E_{\rm bin}| >1000 kT$ binaries as a function of
the distance to the cluster center at an age of 600\,Myr.  These data
combine models W4 and W6.  To guide the eye we have placed a straight
(dotted) line through these bullet points.}
\label{fig:W46_rfbin}
\end{figure}

Figure \ref{fig:W46_rfbin} shows the binary fraction as function of
distance to the cluster center at birth and at $t=600$\,Myr, adopted
as a representative snapshot of an older cluster.  (Qualitatively
similar results would have been obtained had we selected any other
time between $\sim 200$\,Myr and 1\,Gyr.)  The upper solid line shows
the initial distribution for all binaries, regardless of their binding
energy. The lower solid line presents the initial fraction of binaries
with a binding energy of at least $|E_{\rm bin}| > 1000 kT$.  To avoid
clutter, we have shown the statistical uncertainties in the form of
error bars only for the bullet points; the uncertainties in the open
and close triangles can be estimated by the scatter between
neighboring points.

The triangles in Figure \ref{fig:W46_rfbin} shows the distribution for
all binaries at $t=600$\,Myr for model W6 (open) and for model W4
(filled).  The two distributions show similar trends, except for a
generally smaller binary fraction in model W4 (see also Figure
\ref{fig:Bcircfrac}).  The binary fraction in the central portion of
the cluster is considerably higher than it was initially (upper solid
line), because of mass segregation.  For higher $r$ values, the
fraction of objects that are binary remains below the original value,
all the way to and beyond the tidal radius.

The bullets indicate the distribution of binaries with $|E_{\rm bin}|
> 1000 kT$ at a cluster age of 600\,Myr.  In other to improve our
statistics, we have combined the data from the two models; the
distributions are similar, with model W6 containing $\sim 8$\,\% more
hard binaries than model W4.  Note that we find a higher proportion of
hard binaries over the entire cluster field as compared to the initial
distribution.  This is mainly a result of dynamical evolution: hard
binaries tend to get harder, and therefore some of the binaries
contributing to the bullet points started off with a binding energy
less that $1000 kT$, and thus were not counted in the construction of
the lower solid line.

\begin{figure}
\psfig{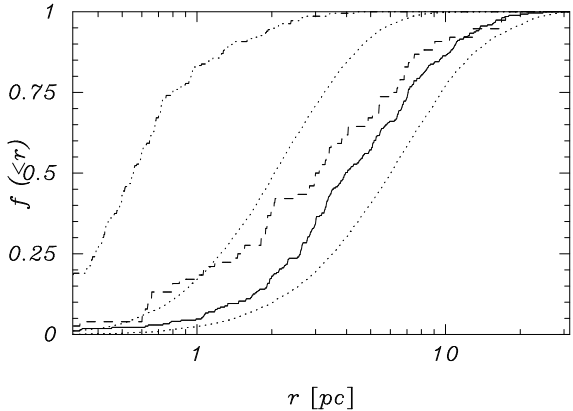} 
\caption[]{Cumulative radial distribution of single stars and binaries
in models W4 and W6 (combined).  The cumulative distribution for all
single stars and binaries at birth and at $t=600$\,Myr are given by
the left and right dotted curves, respectively.  The solid curve gives
the distribution for binaries with $|E_{\rm bin}| > 10^3 kT$ at an age
of 600\,Myr.  The dashed curve gives the same distribution for
$|E_{\rm bin}| > 10^5 kT$ binaries.  The dash-3-dotted curve gives the
radial distribution for higher order (mostly triple) systems.  To
improve statistics we accumulated all triples for models W4 and W6
over the time range of 550\,Myr to 650\,Myr.}
\label{fig:ncumd_r_W4W6}
\end{figure}

The higher central concentration of very hard binaries, as compared to
less hard binaries, is evident in Figure \ref{fig:ncumd_r_W4W6}, which
figure shows the cumulative distribution for single stars, hard
$|E_{\rm bin}| >1000 kT$ and super hard $|E_{\rm bin}| >10^5 kT$
binaries.  The figure also presents the cumulative radial distribution
for higher order (mostly triple) systems.  The left-most and
right-most dotted curves in Figure \ref{fig:ncumd_r_W4W6} show the
radial distribution for model W6 at birth and at an age of 600\,Myr,
respectively.  Initially, both single stars and binaries follow the
left-most dotted curve.  At an age of 600\,Myr, all binaries are more
centrally concentrated than the single stars (see the figure caption).
The harder the binaries, the more centrally concentrated their
distribution has become.  In the case of triples, the condensation
toward the center is extreme, with a triple half-mass radius of around
0.3 pc.

The high central concentration of the very hard $|E_{\rm bin}| > 10^5
kT$ binaries is in part due to their early history.  These binaries
are largely the product of an unstable phase of mass transfer.  The
stellar components in these binaries were therefore more massive than
they are at $t=600$\,Myr, so they are more more strongly affected by
mass segregation, causing them to sink to the central regions.  In
contrast, the moderately hard binaries with binding energies between
$100 kT$ and $10 kT$ (not shown in Figure \ref{fig:ncumd_r_W4W6}) are
barely more centrally concentrated than the single stars.  The triples
are strongly centrally concentrated.  This is a the most dramatic
result of dynamical interactions: triple systems were not initially
present, and must be created dynamically through encounters. Since the
encounter rate increases sharply toward the cluster center, triples
and higher-order systems are born, and often quickly destroyed again,
near the very center of the cluster.

\subsection{Escapers}

\begin{figure}
\psfig{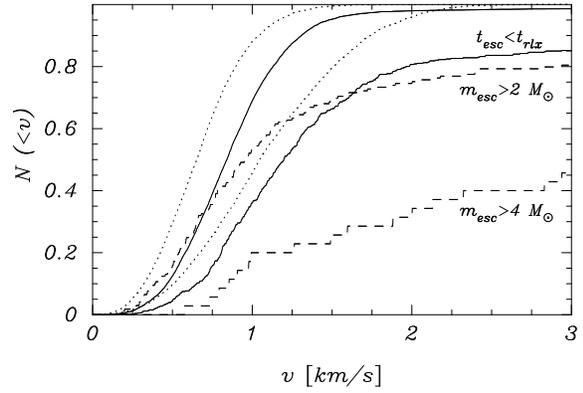} 
\caption[]{Cumulative distribution of the velocities of escaping
single stars and binaries in models W4 and model W6 (combined).  The
two dotted lines show the velocity distributions of single stars in
the models at zero age (right) and at $t=600$\,Myr (left).  The left
solid curve presents the distribution of the escaper velocity of all
stars and binaries integrated over time.  The right solid
curve gives the velocity distribution of stars which escaped within
the first half-mass relaxation time ($\sim 150$\,Myr).  The two dashed
curves give the escape speeds for stars with masses $m_{\rm esc} >
2$\,\msun\, (upper) and $m_{\rm esc} > 4$\,\msun\, (lower).}
\label{fig:W64_vesc}
\end{figure}

The most important mechanism by which stars escape from the cluster is
tidal stripping.  Stars are assumed to have been tidally stripped once
they reach two Jacobi (tidal) radii from the cluster center; they are
then removed from the simulation.  Stripped stars generally leave the
cluster with relatively low velocities, as illustrated in Figure
\ref{fig:W64_vesc}, which shows the distribution of escaper speeds for
models W4 and W6.

It is convenient to draw a distinction between tidal evaporation and
dynamical ejection of stars.  Operationally, we distinguish between
the two processes by the speed of the escaper; an ``ejected'' single
star or binary leaves the cluster with a speed exceeding the escape
velocity. That is, we infer the dynamics from the speed. Typically,
this velocity is imparted to the escaping object during a strong
interaction with another object in the cluster core.  Because of the
relatively low central densities (and hence low reaction rates) in our
model clusters, this type of ejection is rather uncommon.  White dwarf
ejection velocities are not significantly different from those of main
sequence stars, but giants have on average somewhat lower escaper
velocities. This is mainly a consequence of the fact that giants are
generally large and therefore cannot have very close encounters
without colliding, so they are less able to pick up large escape
velocities.

\begin{figure}
\psfig{figure=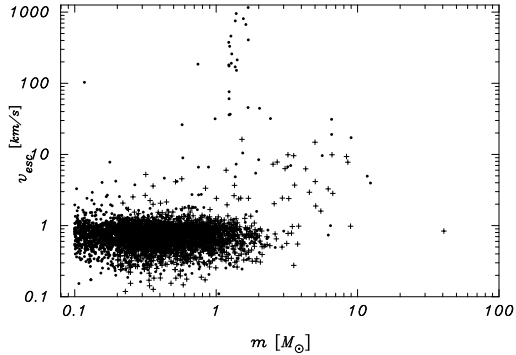,width=7.5cm,angle=-90} 
\caption[]{Mass versus velocity for escaping single stars (dots) and
binaries (plus signs) in model W6.  For the binaries, the total mass
is used.}
\label{fig:W64_Mvesc}
\end{figure}

Another important escape mechanism is supernova explosions.  A neutron
star formed in a core-collapse supernova is hurled into space with a
``kick'' velocity that can easily exceed several hundreds of
kilometers per second, greater than the cluster escape speed by 1--2
order of magnitude.  These objects are hidden in the high-velocity
tail of Figure \ref{fig:W64_vesc}; however, they are clearly visible
in Figure \ref{fig:W64_Mvesc} as the population of high-speed single
objects with masses between 1\,\msun\, and 2\,\msun.

A supernova in a binary system has several effects on the velocity
distribution of both single and binary stars.  The sudden mass loss in
the supernova event, as well as the asymmetric velocity kick, affect a
binary's internal orbital parameters and its center of mass velocity.
When the binary is dissociated in the supernova, the companion to the
exploding star is ejected with its orbital velocity, while the newly
formed compact object receives an additional velocity kick.  The
effects of supernov\ae\, on binary systems, and on the velocities of
the binaries and single stars which result, are reviewed by Portegies
Zwart (2000).

Figure \ref{fig:W64_vesc} shows the velocity distribution of escaping
stars and binaries from models W4 and W6.  Due to the similarities in
escape speeds between the two models, we have combined the data in a
single plot.  The escaper velocity distribution for single stars is
slightly higher than that of the binaries. For clarity we show only
the combined distribution of single and binary escapers in Figure
\ref{fig:W64_vesc}.  Escaper velocities are generally only slightly
higher than the cluster velocity dispersion, and somewhat higher in
the W4 models than in W6.  The small fraction of high-velocity
escapers in Figure \ref{fig:W64_vesc} is due to neutron stars escaping
after supernovae explosions.

In addition to the overall escaper velocities for the combined models,
Figure \ref{fig:W64_vesc} also shows the distribution for stars which
escaped within the first half-mass relaxation time (right solid line).
This distribution has a considerably higher mean than the overall
distribution.  The figure also shows the velocity distributions for
escaping stars more massive than 2\,\msun\, (top dashed curve) and
$m_{\rm esc} > 4$\,\msun\, (lower dashed).  These distributions are
very different from the average escaper speed.  Both models W4 and W6
experience high-density phases during their early evolution, within 1
initial half-mass relaxation time.  The early escapers therefore have
higher velocities.  A similar process operates for the more massive
escapers.  The turn-off mass of a 2\,{\msun} star is about 800\,Myr,
while a 4\,{\msun} star lives for less than about 140\,Myr.  The
distribution of escaper speeds for stars which escaped within {\trlx}
includes the stars with $m>4$\,{\msun}.  Still, the distributions are
considerably different, in the sense that the massive stars have
relatively high escape speeds.  The reason for this discrepancy is the
greater dynamical activity of the high-mass stars, which take part
much more often in dynamical encounters with binaries.

The dependence of escaper velocity on escaper mass is illustrated in
Figure \ref{fig:W64_Mvesc}.  Most escapers have low masses and low
velocities.  The average escaper velocity in the W4 and W6 runs was
1.14\,\kms\, and 1.00\,kms, respectively.  Higher mass stars (and
binaries) have considerably higher velocities. This trend was first
observed by Blaauw (1961,\nocite{blaauw_1961} see also Gies \& Bolton
1986).\nocite{1982ApJ...260..240G} Binaries are mostly ejected by
dynamical encounters.  Supernovae are also responsible for numerous
high velocity escapers, both from kicks and from binary effects, as
mentioned above.  The neutron stars are clearly visible in Figure
\ref{fig:W64_Mvesc} as the objects having masses of about 1.4\,\msun
and velocities up to $\sim$1000\,\kms.

Two rather low-mass ($m\sim 0.11\,\msun$ and m$\sim 0.7\,\msun$) stars
have unusually large ($\sim100\kms$ and $185\kms$, respectively)
escape velocities. The more massive of the two was ejected from a
tight binary at about $t=5.6$\,Myr when its companion exploded in a
Type Ib supernova, dissociating the binary.  The low-mass object was
ejected following a three-body encounter at $t=223$\,Myr.

Of all stars ejected from models W4 and W6, only 109 ($\sim$0.93\% out
of a total of roughly 12,000) have velocities exceeding three times
the dispersion velocity $v_{\rm disp}$ of the parent cluster.  Of
these, 46 are neutron stars.  However, 57\% of escapers with masses
between 4\,{\msun} and 11\,{\msun}, and 75\% of escapers more massive
than 11\,{\msun}, have velocities more than $3v_{\rm disp}$.  The
average escaper velocity of stars between 4\,{\msun} and 11\,{\msun}
is 8.5\,{\kms}; stars with masses exceeding 11\,{\msun} have an
average escaper velocity of 21\,{\kms}, which is much higher than the
cluster dispersion velocity.

These numbers are unusually high compared to the number of runaways in
the Galactic disk.  However, as our criterion here we have adopted
three times the cluster velocity dispersion, instead of Blaauw's
(1961) criterion of 40\,{\kms} (or three times the velocity
dispersion in the Galactic disk).  The fraction of high velocity
stars following Blaauw (1961) was between $\sim 1$\,\% for stars with
$m\aplt 11$\,{\msun}, about 2.5\% for early B stars and about 20\%
among O stars ($m\apgt 16$\,\msun, see also Sone
1991).\nocite{1991AJ....102..333S} Among stars of spectral type A, Stone
(1991) found a runaway frequency of $\aplt 0.3$\,\%. These numbers are
consistent with binary population synthesis studies of binaries in
which stellar dynamical encounters are not taken into account
(Portegies Zwart 2001).


The velocity dispersion of all stars in the cluster at an age of
600\,Myr is $v_{\rm disp} = 0.72\pm0.23$\,\kms. For stars within 1\,pc
(about the core radius) of the cluster center, $v_{\rm disp} =
0.98\pm0.34$\,\kms; stars between 10 and 15\,pc of the center have
$v_{\rm disp} = 0.59\pm0.16$\,\kms.  Most striking is the higher
velocity dispersion for stars well outside the tidal radius
($r>15$\,pc) which have $v_{\rm disp} = 0.75\pm0.20$\,\kms.  A KS test
indicates that the various velocity distributions are significantly
different.

Drukier et al.~(1998)\nocite{1998AJ....115..708D} reported a similar
effect in the Globular cluster M15, and attributed it to tidal
shocking. However, this process is not included in our models; we
therefore conclude that the observed increase in the velocity
dispersion outside the tidal radius cannot simply be a consequence of
tidal shocks.

\subsection{Collisions and Coalescence}\label{Sect:Collisions}

Stellar mergers result from unstable mass transfer in a binary system
or from direct collisions between stars.  Sometimes (unstable) mass
transfer is initiated by the presence of a third star.

\begin{figure*}
\psfig{figure=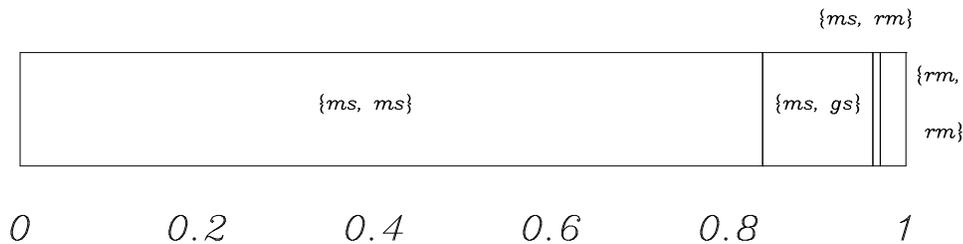,width=14.0cm,angle=-90} 
\caption[]{Schematic overview of the relative frequencies of mergers
between various stellar types in models W4 and W6 (combined).  Stellar
types are denoted by $ms$ for main sequence, $gs$ for (sub)giants and
$rm$ for white dwarfs and neutron stars.  Table \ref{Tab:W6_bin} gives
the numbers of occurrences.}
\label{fig:W46_coll}
\end{figure*}

Table \ref{Tab:W6_bin} gives an overview of the 241 collisions
occurring during the model calculations discussed here.  A schematic
diagram of the relative frequencies of various collision
configurations in presented in Figure \ref{fig:W46_coll}.  This figure
may be compared directly with Figure 2 of Paper\,I, and with Figure 5
of Paper II.  For easier comparison, we have added to Table
\ref{Tab:W6_bin} two columns summarizing the results of Papers I and
II. Note, however, that the comparison with papers I and II is not
really appropriate, as these model calculations lasted for 10\,Gyr.
According to a KS test there is no significant difference between the
time histories of the collisions in the W4 and the W6 models.

\begin{table}
\caption{Mergers occurring in models W4 and W6, compared with those in
model calculations published previously.  The first column identifies
the two stars involved in the merger.  The next two columns give the
number of mergers in each model calculation, followed by the fraction
of each merger type (expressed as a percentage, combining all runs).
The last two columns give the fractions of mergers which occurred in
models S of Papers I and II.  Mergers between two remnants in Paper II
(model S) are included here under \{wd, wd\}, although a few actually
involved a neutron star or black hole.
}
\begin{tabular}{l|rlrrr} \hline
              & W4   & W6   & total &  Paper IS & Paper IIS \\
Ntot:	      & 117  &  124
	      & \multicolumn{3}{c}{------------~~~[\%]~~~------------} \\ 
\hline  
\{ms, ms\}    &  92  &  110 & 84  &  77     &  80 \\
\{ms, gs\}    &  19  &   11 & 12  &  14     &   6 \\
\{ms, wd\}    &   0  &    1 &  0  &   5     &  11 \\
\{ms, ns\}    &   1  &    0 &  0  &   1     &   1 \\
\{gs, ns\}    &   0  &    0 &  0  &   3     &   0 \\
\{wd, wd\}    &   5  &    0 &  2  &   0     &   2  \\
\{wd, ns\}    &   0  &    2 &  1  &   0     &   0 \\
\hline
\end{tabular}
\label{Tab:W6_bin} 
\end{table}

Most ($\apgt 80$\%) mergers in models W4 and W6 occur between two main
sequence stars (see Table \ref{Tab:W6_bin}).  This was also the case
in Papers I and II.  In Paper I, however, only collisions between
single stars were considered, and the high proportion of main-sequence
collisions simply reflects the high proportion of main-sequence stars.
Paper II included both collisions between single stars and mergers
resulting from binary mass transfer, so the results in that case
should compare better with models W4 and W6.  The collisions reported
in this paper, however, have a rather small contribution from \{ms,
wd\} mergers compared to those in Papers I and II.  This may be due in
part to the smaller ages ($\aplt 1$\,Gyr) of our models compared with
Papers I and II ($< 10$\,Gyr).

Although clearly limited by small number statistics, an apparent
difference between models W4 and W6 is the number of \{wd, wd\}
mergers, which are much more common in model W4.  This difference is
reflected in the higher Type Ia supernova rate in model W4.
Interestingly, the calculations in Paper II also had a high \{wd, wd\}
merger fraction.  In fact, if the binary population of models W4 and
W6 were allowed to evolve after the disruption of the cluster, the
rate of \{wd, wd\} mergers would be very similar to the results of
Paper II. The cause of the high proportion of white-dwarf mergers in
model W4 and Paper II is the larger amount of dynamical activity in
these models, which drives more binaries into a state of mass
transfer, favoring the formation of white-dwarf binaries (see also
Hurley \& Shara 2002).  The resulting Type Ia supernova rate is
discussed in the next section.

\begin{figure}
\psfig{figure=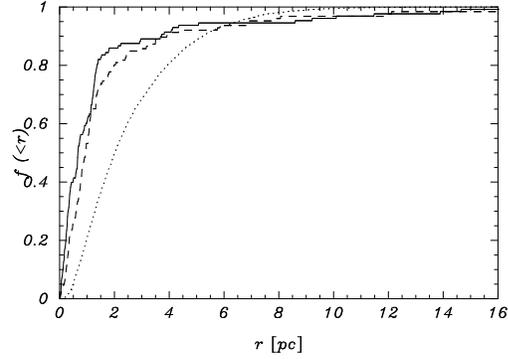,width=7.5cm,angle=-90} 
\caption[]{Cumulative distribution of the distance from the cluster
center at which collisions occur in models W4 (dashes) and W6 (solid).
The dotted line gives the initial distribution of stars in the W6
models. }
\label{fig:W46_rcoll}
\end{figure}

Figure \ref{fig:W46_rcoll} shows the radial distribution of collisions
in models W4 and W6.  More than 80\% of all collisions occur within
the initial half-mass radius ($\sim 3$\,pc).  Collisions in the W6
models tend to be more concentrated to the cluster core than in the W4
models.  It is striking that some ($\sim 5$\%) collisions occur far
($\apgt 8$\,pc) from the cluster center.  These collisions are induced
by binary evolution, whereas mergers in the core are mainly caused by
stellar encounters.

The expected distribution of masses and radii of colliding stars can
be calculated following Portegies Zwart et al.\,(1997; see their
Eq.\,14).\nocite{pzhv97} Assuming a Maxwellian velocity distribution
with velocity dispersion $v$ and including the effects of
gravitational focusing, the number of collisions in the cluster per
$10^8$\,year may be expressed as
\begin{equation}
	\Gamma \approx  \left(\frac{n}
			       {10^3 {\rm pc}^{-3}}\right)^2 
		            \left(\frac{\rcore}{{\rm pc}}\right)^3
			    \left(\frac{ m }{\msun}\right) 
		            \left(\frac{d}{\rsun}\right) 
		            \left(\frac{\kms}{ v }\right),
\label{Eq:rate}
\end{equation}
where $m$ is the stellar mass.  This rate is averaged over all other
masses.  (Note that, based on this estimate, we expect no collisions
during the time span of our simulations for our adopted cluster
parameters, underscoring the importance of binary interactions.)

\begin{figure}
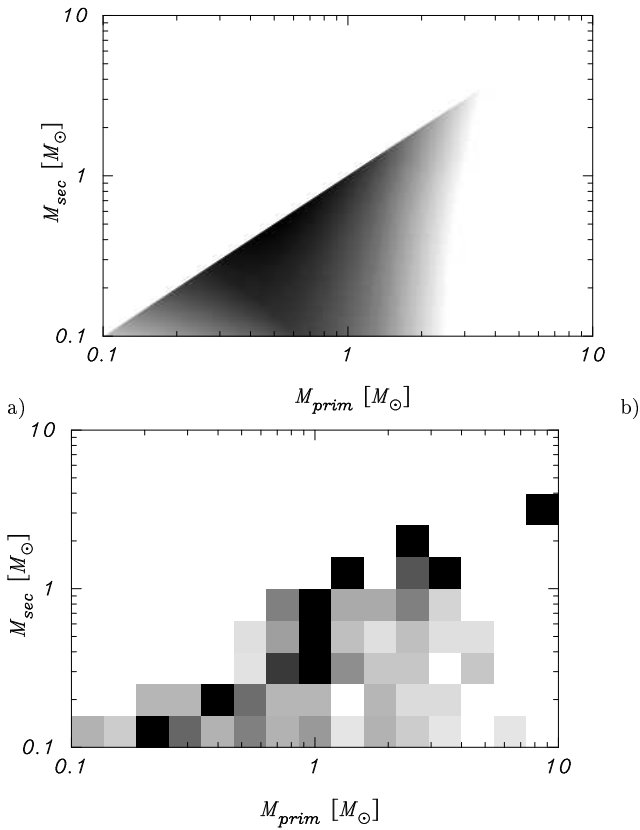

a) \psfig{figure=./fig_W6_Mcoll_expected.ps,width=7.5cm,angle=-90} 
b) \psfig{figure=./fig_W46_Mcoll.ps,width=7.5cm,angle=-90} 
\caption[]{Relative collision rates between a primary and a secondary
star as expected from Eq.\,\ref{Eq:rate} (upper panel), and as found
in models W4 and W6 (lower panel).  For the lower panel the mergers
which resulted from an unstable phase of mass transfer are excluded;
only the true collisions are taken into account.  The shading is
linear in the encounter probability in the upper panel and in the
number of collisions in the lower panel.  Darker shades indicate
higher collision frequency.  Since the probability distribution is
symmetric about the line of equal masses, only the lower half of each
figure is displayed.}
\label{fig:W46_Mcoll}
\end{figure}

Figure \ref{fig:W46_Mcoll} shows the expected (upper panel; see paper
III for details) and observed (lower panel) distributions of primary
and secondary masses of stars involved in collisions in models W4 and
W6. We excluded here the stars which merged as a result of unstable
mass transfer. The upper panel is created via Eq.\,\ref{Eq:rate} from
the initial mass function and the same mass-radius relation for
zero-age main-sequence stars as was used in the model calculations.
Gray shades indicate collision probability; darker shades correspond
to higher values.  The collisions observed in our models are presented
in the lower panel.  The masses of the colliding components in our
model calculations are on average somewhat higher than expected.  This
phenomenon was first observed by Portegies Zwart et al.\, (1999; see
also Portegies Zwart \& McMillan
2002)\nocite{pzmmh99}\nocite{2002ApJ...576..899P} in simulations of
young star clusters having high central stellar densities.  Portegies
Zwart et al (1999) used R\,136, a compact and very dense star cluster
in the 30\,Doradus region, as template for their simulations.  We
return to this comparison in the discussion section.

\begin{figure}
\psfig{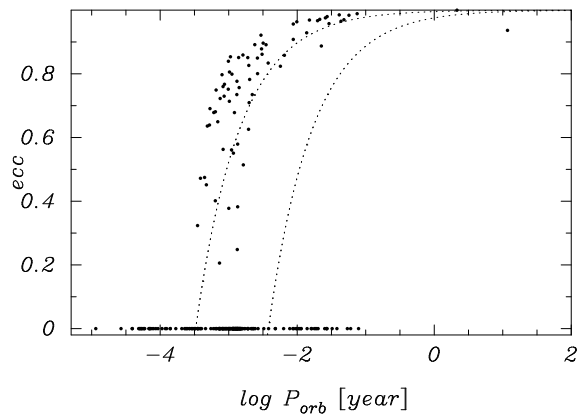} 
\caption[]{Orbital periods and eccentricities of coalescing binaries
in models W4 and W6. Notice the rather large population of circular
(zero eccentricity) binaries which experience a collision.  }
\label{fig:W46_Pecoll}
\end{figure}

For 240 of the 241 stellar pairs experiencing a collision, we were
able to reconstruct the orbit before the collision occurred. Figure
\ref{fig:W46_Pecoll} shows the orbital parameters for the bound
systems thus obtained.  The single pair for which this reconstruction
was not possible experienced a collision during a supernova
event---the neutron star formed in the supernova was ejected directly
into its 1.9\,{\msun} main-sequence companion.  In 181 of the 240
collisions ($\sim 71$\%), the orbit was circular on merger.  The
eccentricity distribution in the other bound cases is consistent with
a thermal distribution.  This result is consistent with the findings
in Paper II, where for model $S$, 19\% of the collisions were the
result of a dynamical encounter.

\subsection{Supernovae}\label{Sect:supernocae}

Supernovae can dramatically affect the evolution of a star cluster.
During a supernova, the exploding star ejects a large fraction of its
mass at high speed.  This mass quickly leaves the cluster, as the
ejection velocity far exceeds the cluster's escape speed.  In
addition, the newborn compact object may also receive a kick velocity
large enough to escape the cluster.  Binaries are very fragile to
supernovae, and each binary hosting a supernova is likely to be
dissociated.

\subsubsection{Supernova types}

Tables \ref{Tab:supernovae} and \ref{Tab:W46_supernovae} present
overviews of the supernovae observed in the models discussed in this
paper, and compare them with expected supernova rates.  Table
\ref{Tab:supernovae} lists supernovae by type; Table
\ref{Tab:W46_supernovae} gives the evolutionary state of the exploding
star and also indicates the possible presence of a binary companion.
For comparison, we also present the expected number of supernovae
given the same initial conditions, but ignoring the effects of
dynamical evolution.

\begin{table}
\caption{Observed supernova types (Obs, from Cappellaro et al.\, 1997)
and supernovae occurring in a standard Scalo (1986) mass function
(model IMF), the population synthesis calculations of model AK by
Portegies Zwart \& Verbunt (1996, model PZV-AK) and in models W4 and
W6, and when the dynamical evolution of the stellar system is ignored
(models W4-nd and W6-nd).  The first column indicates the model,
followed by the number of supernovae of types Ia, Ib, Ic and II. All
numbers are normalized to 1.  }
\begin{tabular}{l|lrlrl} \hline
model	&	Ia&	Ib&	Ic&	II \\
\hline
Obs     &     0.09&   0.11&   0.18&   0.62 \\ 
IMF	&      ---&    ---&   0.12&   0.88 \\
PZV-AK	&     0.03&   0.13&   0.12&   0.72 \\
W4	&     0.22&   0.20&   0.07&   0.50 \\
W4-nd	&     0.06&   0.20&   0.17&   0.57 \\
W6	&     0.10&   0.20&   0.10&   0.63 \\
W6-nd	&     0.04&   0.23&   0.08&   0.65 \\
\hline
\end{tabular}
\label{Tab:supernovae} 
\end{table}

\begin{table}
\caption{Overview of the core-collapse supernovae which occurring in
models W4 and W6.  The progenitors are presented in the first column.
When two stars are enclosed by parentheses, the left star explodes.
Two stars enclosed by braces are the result of a merger or collision.
The second column identifies the supernova type.  Subsequent columns
indicate how many core-collapse supernovae occurred in the models.
The subdivisions {\em dyn} and {\em non-dyn} refer to models with and
without dynamics.  Superscript numbers indicate the numbers of
binaries that survive the supernova.  The rows below $N_{\rm esc}$
indicate the numbers of escapers experiencing a core collapse
supernova {\em after} being ejected from the cluster.  The table does
not list the 3 stars in each model that collapse into black holes.
The $^\star$ indicates that one supernova occurred after a phase of
mass transfer to a neutron star.
}
\begin{tabular}{l|lrlrl} \hline
             &Type & \multicolumn{2}{c}{W4}  & \multicolumn{2}{c}{W6} \\ \hline
             &Type & dyn  & non-dyn & dyn  & non-dyn \\ \hline
Ntot:	     &     & 38   & 35   &  25  & 26   \\ \hline
wr           & Ic  &  1   & 4$^2$&    0 &  2 \\
gs           & II  & 16   &  13  &   12 & 13 \\
he           & Ib  &  1   &  2   &    0 & 2$^1$ \\
(gs, ms)     & II  &  4   &  7   &    6 & 4 \\
(gs, gs)     & II  &  2   &  0   &    0 & 0 \\
(gs, bh)     & II  &  1   &  0   &    0 & 0 \\ 
(he, ms)     & Ic  &  3   &  2   & 2$^1$& 0 \\
(he, ms)     & Ib  & 7$^1$&  5   &    2 & 2$^1$  \\ 
(he, he)     & Ib  &  1   &  0   &    1 & 2  \\
\{wd, wd\}   & Ia  &  1   &  2   &    1 & 1 \\ 
\{gs, gs\}   & II  &  1   &  0   &    0 & 0 \\ 
\hline
\hline
$N_{\rm esc}$&total& 16   &      & 6   \\
\hline
gs           & II  &  2   &      & 1   \\
\{wd, wd\}   & Ia  & 11   &      & 2   \\
(gs/he, ms)  &Ib/II&  2   &      & 3$^\star$\\
\{ms, ms\}   & II  &  1   &      & 0   \\
\end{tabular}
\label{Tab:W46_supernovae} 
\end{table}

The adopted Scalo (1986) initial mass function contains 0.41 stars per
thousand with $m>25$\,\msun\, and 3.1 stars per thousand with
$8<m<25$\,\msun.  In a naive evolution model, single stars with masses
$\apgt 25$\,\msun\, may result in a Type Ic supernova, while stars
more massive than 8\,\msun\, produce a Type II supernova.  For a
population of 2000 single stars we then expect $\sim 0.82$ Type Ic and
$\sim 6.2$ Type II supernovae.  (Note that this estimate does not
include binaries, and is therefore inapplicable to Type Ia and Ib
supernovae.)

We compare these numbers with purely binary evolution models of
Portegies Zwart \& Verbunt (1996). According their model AK, the Type
Ic supernova rate is enhanced compared to models of only single stars.
In the non-dynamical models (W4-nd and W6-nd), Type Ia and Ib
supernovae occur without affecting the other supernova types.  One
reason for this is the fact that many Type Ib supernova originate from
stars that were once part of a mass-transferring binary.  The primary
star in these binaries tends to lose its envelope in a phase of mass
transfer and explode in a Type Ib supernova.  The companion star,
although initially not massive enough to experience a supernova, may
accrete some material from the primary star.  This accreted material
may be sufficient to raise the secondary star's mass over the limit
for experiencing a supernova.  Stellar mass is, in a sense, recycled
to produce more supernovae (see also Portegies Zwart \& Yungelson
1999).\nocite{1999MNRAS.309...26P}


The relative numbers of supernova types Ia:Ib/c:II for the 8 model
clusters (models W4 and W6 combined) are 0.18:0.27:0.55; in the models
where stellar dynamics is ignored these ratio are 0.05:0.34:0.61.  The
addition of dynamical interactions to the models enhances the Type Ia
supernova rate at the expense of the Type Ib/c and Type II supernova
rates.  The enhancement of Type-Ib/c supernova in model W4 is due to
close (post mass transfer) binaries, possibly because these binaries
experience more dynamical encounters during the shallow collapse of
the cluster core (see the discussion in \S\ref{sect:evolution}).

\subsubsection{Binary survival rate}

Only two binaries survive their first supernova.  One remains bound
because the supernova results in a black hole, which does not receive
a velocity kick.  This binary experiences a second phase of mass
transfer (see \S\ref{sect:X-ray} for details).  The other surviving
binary is the result of a supernova in a rather short-period binary
with a main-sequence companion.  The resultant neutron star receives
only a mild kick and the binary remains bound.  Later this binary
becomes an X-ray source (see \S\ref{sect:X-ray} for details).

Two supernovae are triggered by collisions between carbon stars,
resulting in Type Ia events, and one Type II supernova follows a
collision between two supergiants.  One neutron star is shot into its
1.9\,{\msun} main sequence companion following a supernova.  The
resulting collision product, a Thorn-Zytkow star, is ejected from the
star cluster before collapsing to a black hole.

Only two out of 29 binaries survive the formation of a neutron star
and one survives the formation of a black hole.  A total of six black
holes result from type Ic supernovae.

\begin{figure*}
\psfig{figure=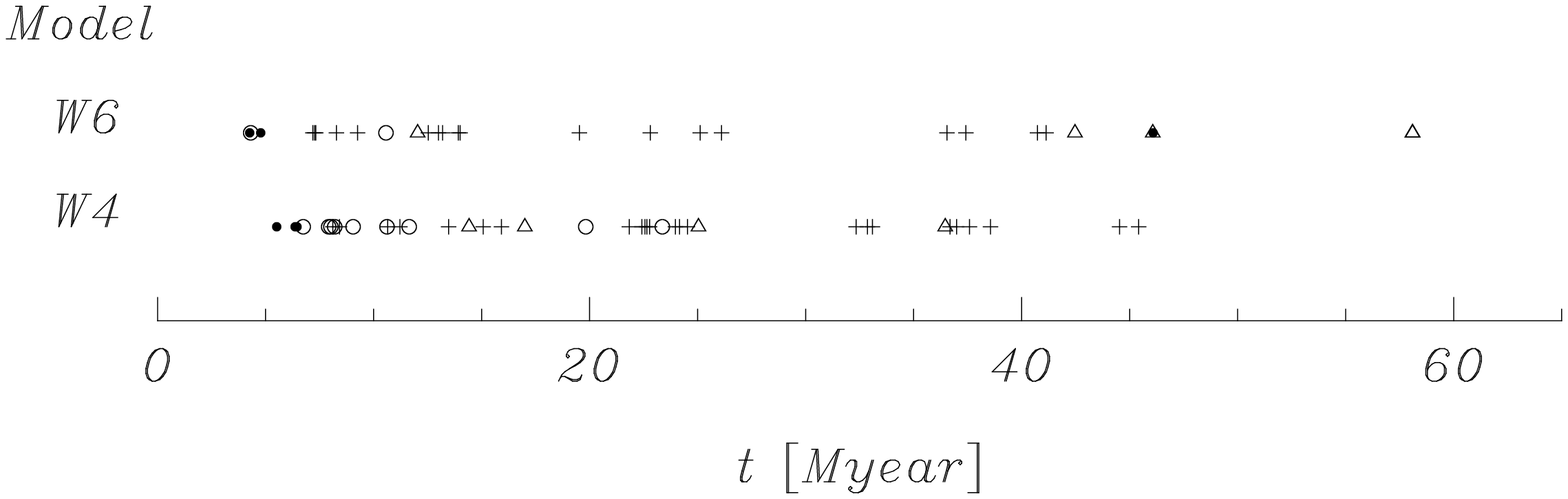,width=14.cm,angle=-90} 
\caption[]{Supernova history for models W4 (bottom) and W6 (top).
Type Ib, Ic and II supernovae are identified with circles, triangles,
and plus-signs, respectively.  Black-hole formation is identified with
$\bullet$.  Note the collapse of a neutron star to a black hole in one
of the W6 models around $t\sim 46$\,Myr (filled triangle).  The time
histories of these two distributions are not significantly different.}
\label{fig:W46_tsupernova}
\end{figure*}

Figure \ref{fig:W46_tsupernova} gives the time history of the
supernovae in models W4 and W6.  First black holes (bullets) are
formed, followed by neutron stars (circles, triangles and plus signs).
The supernovae in the W4 models seem to occur somewhat earlier than in
the W6 models, but the mean time at which a supernova occurs,
23.3\,Myr and 23.5\,Myr for models W4 and W6, respectively, are not
significantly different.  The formation of a black hole at 46\,Myr in
model W6 is the result of a radio pulsar colliding with its carbon-star
companion (see \S\ref{sect:X-ray}). 

\subsection{Mass transfer and peculiar binaries}\label{sect:X-ray}

\subsubsection{First Roche-lobe contact}

\begin{figure*}
\psfig{figure=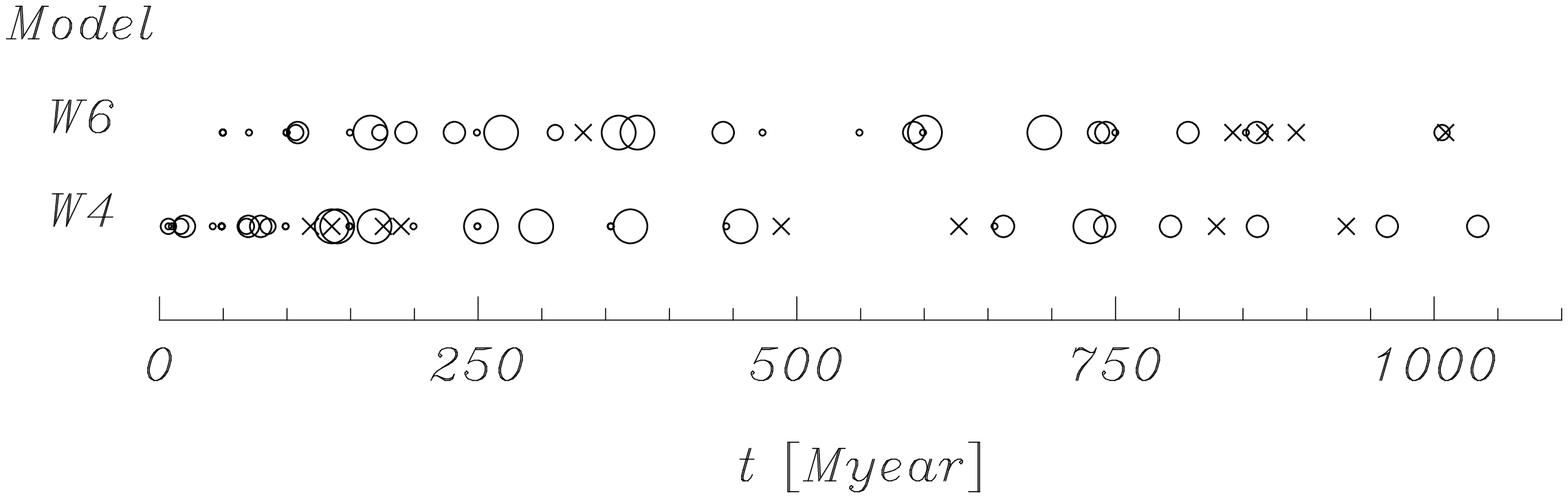,width=15cm,angle=-90} 
\caption[]{Times of first mass transfer for binaries in models W4 and
W6.  The circle size indicates the evolutionary state of the donor,
from main-sequence (small) via Hertzsprung gap, and sub-giant, with
supergiant indicated by the largest circles.  The $\times$ symbols
indicate mass transfer onto a white dwarf (or a black hole, for the
second $\times$ from the right in model W4). }
\label{fig:W46_first}
\end{figure*}

The high binary fractions in our models result in frequent episodes of
mass transfer.  This is illustrated in Figure \ref{fig:W46_first},
which shows the distribution of the times of first Roche-lobe contact
in models W4 and W6.  The various symbols indicate the state of the
donor at the moment of first contact.  (Note that helium stars,
although likely to be the remnant of an earlier phase of mass
transfer, are still counted in this figure.).

In most cases the mass is transferred to a main-sequence star, except
in the cases indicated with $\times$, where the mass transfer is onto
a compact object (mostly a white dwarf).  In the majority of these
latter cases the donor is a (sub)giant, but there are two occasions
where the donor is a helium star, and in one case a main-sequence star
(see Figure \ref{fig:W46_cv}.  One episode of mass transfer occurs
between a (sub) giant and a black hole (model W4 at $t=829$\,Myr).

The time histories of models W4 are significantly different from those
of models W6, according to a $\chi^2$ test on the cumulative
distributions of Figure \ref{fig:W46_first}. In the W4 models, mass
transfer tends to occur earlier than in the W6 models.  As discussed
above, the difference is a result of the greater dynamical activity
during the early phase of model W4.  In fact, the higher binary
activity in this case is caused by the early phase of shallow core
collapse in the W4 models, which does not occur in the W6 models.

\subsubsection{X-ray binaries}

\begin{figure*}
\psfig{figure=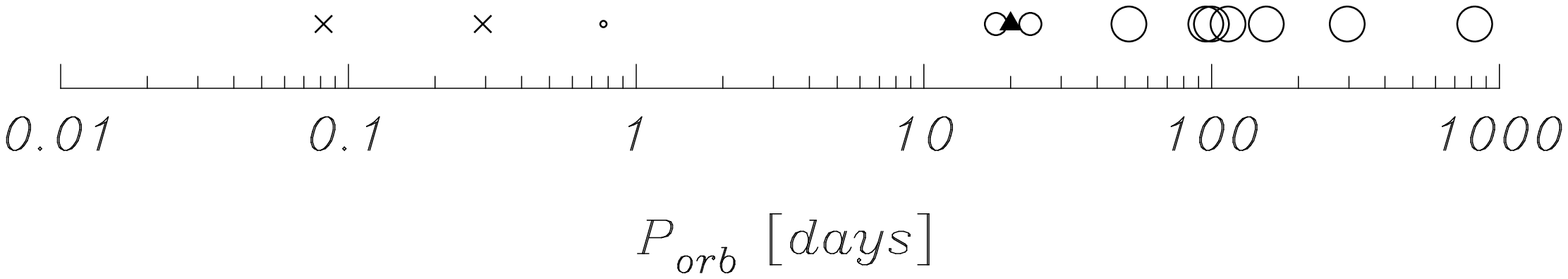,width=15cm,angle=-90} 
\caption[]{Orbital period (in days) at the onset of mass transfer onto
a compact object, for models W4 and W6.  The circle size indicate the
evolutionary state of the donor, from main sequence (smallest circles)
to supergiant (large circles).  The $\times$ indicates that the donor
is a helium star; the filled triangle indicates the period of the
simgle black hole X-ray binary.}
\label{fig:W46_cv}
\end{figure*}

Mass transfer from a hydrogen- or helium-burning star onto a compact
object generally leads to an X-ray phase.  Figure \ref{fig:W46_cv}
shows the distribution of orbital periods (in days) at the onset of
Roche-lobe overflow for binaries with white-dwarf (in one case a black
hole) accretor.  Note the clear distinction between main-sequence and
helium-star donors at short orbital periods and giant donors at larger
periods.  The wide systems with giant donors are probably most
comparable to the supersoft sources (X-ray binaries where a white
dwarf accretes at a super-Eddington rate for a solar-mass object,
emitting at a rate of about $10^{38}$ erg/s).

A peculiar object which might be observable as an X-ray source results
from a binary in which a helium giant (2.5\,\msun) transfers mass to a
1.17\,{\msun} carbon-oxygen white dwarf.  First contact occurs at
45.9\,Myr at an orbital period of 1.7 days.  When the carbon star
explodes and becomes a neutron star (at 46.07\,Myr) it receives a kick
of only 16\,\kms.  At that moment the companion star still fills its
Roche lobe, and as a result the two stars coalesce into a single
object.  The merger product collapses a little later to form a
2.42\,{\msun} black hole.

\subsection{Triples and higher order systems}\label{sect:triples}

Long-lived triples and higher-order systems pose severe challenges to
any numerical code.  We encountered a total of 41 long-lived multiples
in our runs, 21 triples [4 quadruples] in W4 and 13 triples [3
quadruples] in W6.  Only one of the quadruple systems is hierarchical;
the others are binary-binary systems.  

\begin{figure}
\psfig{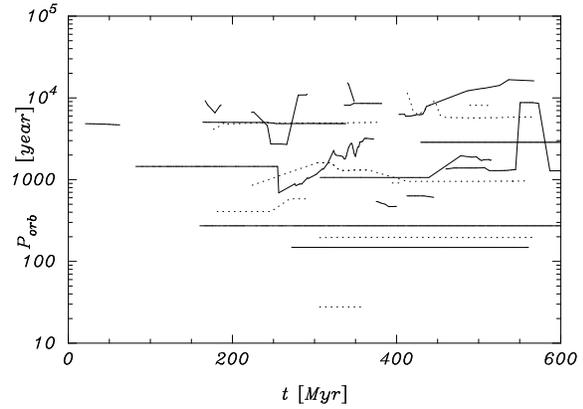} 
\caption[]{Orbital periods of the outer components of all hierarchical
triples in all models, solid curves for the multiples in models W4,
the dotted lines for the W6 multiples.}
\label{fig:W46_tP_triple}
\end{figure}

The first multiple systems form as early as 20\,Myr (by a
binary-binary interaction); some survive for as long as 900\,Myr.
Figure \ref{fig:W46_tP_triple} presents an overview of all triples in
all models from $t=0$ to 600\,Myr.  (At later times the numbers of
triples decrease as the clusters dissolve.)  The orbital period of the
outer binary is generally rather constant. This is not surprising as,
for the triple to be long-lived, it must be hierarchical and isolated,
and the orbital energies and hence periods are adiabatic invariants.
The systems showing a significant period derivative are driven by
binary evolution or by temporary close encounters with other cluster
members.

The fractions of long-lived triple and quadruple systems in our models
are only $\sim 0.26$\% and $\sim 0.12$\%, respectively, considerably
smaller than the fraction of triples observed in the Galactic field
(2.6\%, Duquennoy \& Mayor 1991)\nocite{1991A&A...248..485D} or in the
Pleiades cluster ($\sim 2$\%, Mermilliod et al. 1990,
1992)\nocite{1990A&A...235..114M}\nocite{1992A&A...265..513M}.  Note
also that these are the total numbers of multiple systems created at
any time during our simulations.  The mean lifetime of such a system
is $280\pm180$\,Myr.

We can study the population of short-lived multiple systems by
selecting a bound pair of stars and searching for a third nearby star
which forms a bound subsystem.  At any moment in time the number of
such short-lived multiple systems is about twice as high as the number
of persistent triples.  At $t=600$\,Myr we counted a total of 46 bound
multiple systems in all models (W4 and W6), implying in a triple
frequency of about 0.6\% per star.  For comparison, Kroupa
(1995)\nocite{1995MNRAS.277.1522K} found a triple fraction of 0.5\%
using a similar technique. The outer orbits of these triples, however,
are generally rather soft.


Nine collision products in our models W4 became members of persistent
triples or quadruples. These collisions occured in transient
triples. In about half (5) the cases, the two stars in the inner
binary collided while the intruding star remained bound to the merged
star.  (For definiteness, we take the ``inner'' orbit in a quadruple
system to be the larger of the two inner orbits.)  In a few cases (3),
the merger product and its companion became bound to another third
star, forming a stable triple. In two of these cases, the encounter
resulted in second collision with the merger product. Both these
triple collisions resulted in blue stragglers which had, at some time,
a mass exceeding twice the turnoff mass of the cluster.

Of the total of 117 collisions in the W4 models, nine occured in
transient triple systems.  With a triple frequency of only $\sim
0.26$\%, one naively expects less than one collision to occur in
triples, making the frequency of collisions in triples unusually high.
In part, this higher fraction of collisions in multiples is due to the
dynamics---such systems generally form close to the cluster center,
where the stellar density is highest and most collisions are expected.
In addition, collision products tend to be found near the center,
because they are more massive than average.  The higher combined mass
of three stars and the relatively large cross section of the outer
orbit also increases the collision probability.  Similar circumstances
led to the high collision rates observed in the R\,136 simulations of
Portegies Zwart et al.\,(1999, see also Portegies Zwart \& McMillan
2002). Thus, triples are favorable for collisions, and collision
products favor triple systems.

We can characterize the triple systems formed in our simulations as
follows: All inner orbits have periods less then 100\,year with an
average of about 5.1 years. All outer orbital periods exceed 100\,year
and have an average of about 4600 years. The eccentricities of the
inner and outer orbits are quite distinct. The inner orbits have
generally rather small eccentricies ($e = 0.44\pm 0.3$) and its
distribution is consistent at a 98.8\% level with the inner
eccentricity distribution of observed triples (Tokovinin
1997).\nocite{1997A&AS..124...75T} The distribution of outer
eccentricities ($e = 0.65\pm 0.14$) is consistent with thermal.


\section{Comparison with previous work}

\subsection{McMillan \& Hut 1994}

The simulations reported by McMillan and Hut (1994) used up to 2048
stars, with up to 20\% (rather soft: $1-20$ or $1-100 kT$) primordial
binaries, and included the Galactic tidal field.  However, they
excluded stellar evolution and hence any stellar mass loss.  In the
absence of a physical time scale associated with stellar evolution,
they presented their results in units of the initial relaxation time.
Our W6 models have half lives of about 6 initial relaxation times,
much shorter than the $\sim12$--30 initial relaxation times for the
most comparable McMillan \& Hut models.  Stellar mass loss is the main
reason for the more rapid dissolution of our models; in addition, the
McMillan \& Hut runs all began with the cluster well inside its Jacobi
surface, so the clusters had to expand significantly before
significant tidal mass loss occurred.  As discussed previously, all
the McMillan \& Hut models experienced core collapse, which is not
seen in our simulations, where core collapse is arrested by stellar
mass loss.

McMillan \& Hut found that the binary fraction in their simulations
first fell due to as binaries were destroyed by interactions with
other binaries in the cluster core, then rose again at late times as
the cluster evaporated in the Galactic tidal field.  In contrast, we
find that the (initially $\sim50$\%) binary fraction in our models
remains roughly constant throughout the calculation, then increases
significantly to $\sim 70$\% when only $\sim 10$\% of the cluster mass
remains.  These differences are most likely attributable to the lack
of significant core collapse and the stronger tidal fields in our
runs.

One of the most interesting conclusions made by McMillan \& Hut is
that the spatial distribution of hard binaries is different than the
distribution of soft binaries (see their Figure 9).  In
Figure \ref{fig:W46_rfbin} we showed that hard binaries $E>1000 kT$)
are slightly more centrally concentrated than average.

\subsection{Kroupa 1995}

The main difference between our calculations and those of Kroupa
(1995) is his neglect of binary evolution and his inclusion of a
prescription for pre-main-sequence binaries.  The eccentricity
distribution of his 'primordial' binaries was therefore somewhat
different from ours, but this affects only binaries with the highest
eccentricities and the shortest orbital periods.  On the other hand,
his neglect of binary evolution underestimated the fraction of tidally
circularized binaries.  In his calculations the boundary between hard
and soft binaries was at an orbital period of roughly $2.7\times
10^4$\,years, similar to that in our models W4.  We have already
compared the characteristics of the populations of triple systems in
our calculations (see \S\ref{sect:triples}).

\subsection{De La Fuente Marcos 1997} 

De la Fuente Marcos (1997) performed studies of the evolution of small
($N\le750$), tidally limited open clusters having a variety of initial
mass functions, with and without the inclusion of stellar (but not
binary) evolution.  All models started with a substantial fraction
(1/3) of primordial binaries having mass ratios of 0.5 and energies in
the $\sim1$--$10 kT$ range.  He found that the dissolution time scale
of his models depended quite sensitively on the choice of IMF, and
that the binary population shortly before dissolution could show
characteristic features allowing remnants of rich and poor clusters to
be distinguished observationally.  It is unclear how these results
extend to larger systems.

The binary fraction in the de la Fuente Marcos simulations ranged from
33\% initially to about $51\%\pm0.19$\% near the disruption of the
cluster (averaged over all 20 of his simulations).  By the time the
clusters dissolved, the binary fraction in the core had dropped to
$15\%\pm0.13\%$.  We find a similar increase in the total binary
fraction, but clearly the core binary fraction in our models is much
higher (see Figure \ref{fig:W46_rfbin}).  The small binary fractions
in the core near the end of his simulations are the direct result of
his choice of initial binary binding energies.

\subsection{Papers I, II}

The comparison between the relative collision frequencies in
Tab.\,\ref{Tab:W6_bin} and those of model $S$ in Paper I are quite
striking.  In that paper only encounters between single stars were
studied and primordial binaries were neglected.  Still, in Paper I
77\% of all collisions occurred between two main-sequence stars, and
14\% between a main-sequence star and a giant; the remaining 9\%
involved giants and remnants.

The (ms, ms) merger rates in our models are comparable to the merger
rates in paper II, but (ms, wd) were much greater in model S of paper
II because it is older (10Gyr).  The (ms, gs) merger rate is lower
because there are fractionally fewer main-sequence stars in paper II.

\subsection{Hurley et al.~2001}

Hurley et al.\, (2001) have modeled the open cluster M67, using a
simulation code similar in many ways to our own.  Both codes use
fitting formulae for single star evolution, and include a series of
recipes for dealing with binary star evolution.  The main difference
between Hurley et al.\, (2001) and the current paper lies in the
parameter choices for their simulations: they evolved 15,000 stars to
an age of 2500 Myr, before adding dynamics, in the form of a direct
$N$-body calculation.  They then evolved the combined model to an age
of 4300 Myr, appropriate for the old open cluster M67.  They also
present the results from smaller $N$-body runs, starting at earlier
times.

Hurley et al.\, (2001) focused mostly on the formation of blue
stragglers, and reported that their simulations indeed produced
roughly the right number of blue stragglers, in agreement with
observations of M67.  They also performed a non-dynamical binary
evolution population synthesis, and found that the number of blue
stragglers in that case fell short of that required by observations.
They concluded that dynamical encounters have been crucial in the
evolution of M67.  While it is difficult to make a quantitative
comparison between their simulations and ours, given the rather
different types of initial conditions, our main results concerning the
numbers and types of collisions, and the formation rate of triples are
in broad agreement.



\section{Summary and Conclusions}\label{sect:discussion}

We have performed detailed {\nbody} calculations of intermediate-mass
open clusters near the Sun.  The initial conditions were selected to
mimic star clusters such as Pleiades, Praesepe and Hyades.  Our
calculations included the effects of dynamical encounters between
stars and higher order systems, the tidal field of the parent Galaxy
and the evolution of single stars and binary stars.

\subsection{Cluster lifetime and structure}
Our model star clusters dissolved in the tidal field of the Galaxy
within about 1 billion years.  The rate of mass loss remained roughly
constant, at $\sim0.8$--$1.4\,{\msun}$ per million years,
corresponding to a cluster half-life of $\sim600$--1000\,Myr,
depending on distance from the Galactic center.

The density profiles of the model clusters changed dramatically during
the cluster lifetime.  Core collapse was prevented by stellar mass
loss and binary heating.  Our less concentrated (W4) models became
more compact during the first few million years, whereas the more
concentrated (W6) models expanded from the beginning.  The W4 models
still dissolved more quickly in the Galactic tidal field, however,
mainly due to their closer proximity of the Galactic center.

Mass segregation is a very efficient process.  After only a small
fraction of an initial relaxation time, single white dwarfs,
giants, collision products, hard binaries, mass transferring binaries
and binaries with one or two massive components (compared to the mean
mass in the cluster) were all noticeably more centrally concentrated
than the average star, as measured by the half mass radii of the
various components.

\subsection{Binary stars}
The numbers of binaries decreased with time at roughly the same rate
as the numbers of single stars, with the result that the binary
fractions in our models remained more or less constant over the
studied range of cluster ages.  The binary fraction generally
decreased during the first few million years by several percent, due
to supernovae and dynamical encounters, both of which tended to
disrupt binaries.  For the remainder of the evolution, this fraction
slowly increased.  The maximum binary fraction of $\sim 65\%$ is
reached near disruption.  The fraction of binaries in the core
remained between 50\% and 60\% higher than the fraction near the
half-mass radius over the entire lifetime of the cluster.

The widest binaries were dissociated soon after the start of the
simulations.  A considerable fraction ($4.5\pm0.6\%$) of highly
eccentric binaries with rather short orbital periods circularized and
some experienced mass transfer early in the evolution of the cluster.
These short-period binaries were generally more centrally concentrated
than wider binaries or single stars, because they are on average more
massive than single stars and do not interact.

The overall distributions of binary parameters, however, hardly change
with time.  Apart from the loss of binaries with the largest orbital
periods and those with short orbital periods and high eccentricities,
most binaries were relatively unaffected by either stellar evolution
or stellar dynamics.  The observed binary populations in open clusters
may therefore be a reasonable representation of the initial primordial
population.

\subsection{Escapers}

We define three families of escaping stars.  (1) Neutron stars were
ejected from the cluster with very high velocities.  About 0.4\% of
all stars were ejected as neutron stars due to supernova explosions.
(2) Massive stars tended to have rather high escape velocities,
because they were often found in close binaries which were disrupted
by the explosion of the primary star, possibly after a phase of mass
transfer.  The average escape velocity of all single main sequence
stars in model W4 was $\vesc = 1.2\pm0.51$\,{\kms}, whereas stars with
$m\apgt 4$\,{\msun} had $\vesc = 17.4\pm2.3$\,{\kms}.  For model W6
these numbers are only slightly smaller.  (3) The mean escaper
velocity of low-mass stars ($\vesc = 1.0\pm0.28$\,{\kms} for stars
with $m<0.5$\,\msun) was comparable to the cluster velocity
distribution; the mean velocity of escaping binaries was ($\vesc =
0.97\pm0.29$\,{\kms} for model W4 and $\vesc = 0.81\pm0.29$\,{\kms}
for model W6), somewhat smaller than that for single stars.

\subsection{Stellar collisions}
Not surprisingly, collisions tended to occur near the cluster center,
but a considerable fraction of mergers occured farther out, near the
tidal radius.  More than 80\% of all collisions occurred within about
two core radii.  Most ($\sim 80$\%) collisions occurred between two
main sequence stars.  About 70\% of the collisions were the result of
an unstable phase of mass transfer.  The rest occurred in
single-star--binary or binary--binary encounters.  Multiple collisions
were rare: only two out of 241 in our model calculations.

Collision participants were somewhat more massive than expected based
on the simple cross section arguments presented in Paper III.  Most
(85\%) collisions occurred between main-sequence stars; most of the
remainder involved a main-sequence star and a giant (12\%). Only 3\%
of the collisions occurred between two remnants. The dynamical
activity in the cluster core, however, caused quite a few white dwarfs
to become binary members, which would later merge due to the emission
of gravitational waves. This leads to a rather high merger rate among
white dwarfs (see also Shara \& Hurley
2002).\nocite{2002ApJ...571..830S}

\subsection{Supernovae and stellar remnants}
The overall supernova rate in our simulations was consistent with
rates observed in the Galaxy.  However, the ratios of Type II to Type
Ia, Ib, and Ic supernova are quite different.  Most striking is the
order of magnitude enhancement of type Ia supernovae in the W4 models
compared to the non-dynamical models and the population synthesis of
non-interacting binaries.  In these same models, Type Ib supernova are
enhanced by about a factor of two compared to a population of binaries
without dynamical encounters.

Because of the assumed neutron-star kick distribution, pulsars were
generally not retained in our cluster models.  Only one out of 61
neutron stars was retained to form an X-ray binary later in its
lifetime.  The observed velocity distribution of ejected pulsars was
somewhat smaller than the input kick velocity distribution.  X-ray
binaries in which a white dwarf accretes from a companion star were
about an order of magnitude more common.  X-ray binaries containing
black holes were rare simply because such objects form from the
highest-mass stars, which are themselves rare.

Among the more unusual collisions, in our simulations, we observed one
between a neutron star and a helium giant, two between CO white
dwarfs, resulting in type Ia supernovae and the formation of an X-ray
binary with a black hole as accretor.  Several white dwarfs
experienced a phase of mass transfer from a companion star.  The donor
was most often a (sub)giant.  In total, 13 X-ray binaries were formed,
most with a white dwarf as accretor.

\subsection{Multiple systems}
The number of persistent triples formed in our calculations was about
an order of magnitude smaller than the observed fraction of triples in
open clusters and the Galactic disk, although the frequency of triples
and higher-order systems present at any moment in our simulations was
comparable to the numbers actually observed.  The majority of these
were transient, however, and their orbital parameters differed
considerably from observations.  The orbital parameters of the outer
orbits of the formed triples formed in our models was not
representative of triples observed in the Galaxy, which tend to have
shorter periods and lower eccentricities.



\section*{Acknowledgments}

This work was supported by NASA through Hubble Fellowship grant
HF-01112.01-98A (NASA contract NAS\, 5-26555) by the Space Telescope
Science Institute, the Royal Netherlands Academy of Sciences (KNAW),
the Netherlands organization for scientific research (NWO), by NASA
ATP grants NAG5-6964 and NAG5-9264. Calculations were performed on the
GRAPE-4 systems at Tokyo University, Drexel University and the
University of Bloominton. SPZ is grateful to Drexel University, Tokyo
University and the Institute for Advanced study for their hospitality.

\bsp
\label{lastpage}
\end{document}